\documentclass[lettersize,journal]{IEEEtran} 
\usepackage{amssymb,amsmath}
\usepackage{cite}
\usepackage{tabularx}
\usepackage{graphicx,subfigure}
\usepackage{psfrag}
\usepackage{url}
\usepackage{hyperref}
\usepackage{booktabs}
\usepackage[latin1]{inputenc}
\usepackage{lscape}
\usepackage{multirow}
\usepackage{tablefootnote}
\usepackage[absolute,overlay]{textpos}
\usepackage{xcolor}
\usepackage{colortbl}
\definecolor{lightgray}{gray}{0.95}
\definecolor{lightgray2}{gray}{0.8}
\usepackage{multirow}
\usepackage{makecell}
\usepackage{array}
\newcolumntype{L}[1]{>{\raggedright\let\newline\\\arraybackslash\hspace{0pt}}m{#1}}
\newcolumntype{R}[1]{>{\raggedleft\let\newline\\\arraybackslash\hspace{0pt}}m{#1}}
\usepackage{arydshln}
\usepackage{verbatim}

\begin{document}	
    \title{Power from Space: Coordinated Satellite Charging for Off-Grid Wireless Systems}
    \author{
        \IEEEauthorblockN{
        Osmel M. Rosabal,~\IEEEmembership{Member,~IEEE},
        Amirhossein  Azarbahram,~\IEEEmembership{Graduate Student Member,~IEEE},
        Mateen Ashraf,
        Mohammad Shehab,~\IEEEmembership{Member,~IEEE},
        Abdul Basit Khattak,~\IEEEmembership{Graduate Student Member,~IEEE},
        Onel L. A. L\'opez,~\IEEEmembership{Senior Member,~IEEE}, and Mohamed-Slim Alouini, ~\IEEEmembership{Fellow,~IEEE}
        }
    \thanks{Osmel M. Rosabal, Amirhossein Azarbahram, Mateen Ashraf, Abdul Basit Khattak, and Onel L. A. L\'opez, are with the Centre for Wireless Communications (CWC), University of Oulu, Finland. \{osmel.martinezrosabal, amirhossein.azarbahram, mateen.ashraf, abdul.khattak, onel.alcarazlopez\}@oulu.fi.}
    \thanks{Mohammad Shehab is with the Networks Department, German University in Cairo, New Cairo 11835, Egypt (e-mail: mohammad.shehab@guc.edu.eg).}
    \thanks{Mohamed-Slim Alouini is with CEMSE Division, King Abdullah University of Science and Technology (KAUST), Saudi Arabia. email: slim.alouini@kaust.edu.sa}
    \thanks{This work is partially supported by Research Council of Finland (Grants 348515 (UPRISING) and 369116 (6G Flagship)).}}
    \maketitle
    \begin{abstract}
        Satellite-enabled wireless power transfer (WPT) may be a transformative solution for charging Internet of Things (IoT) devices in off-grid
        scenarios where traditional technologies struggle to efficiently meet urgent energy demands. In this article, we review the advantages and limitations of microwave-based long-distance charging for satellite-enabled WPT. We then introduce our vision of coordinated space-based WPT, where multiple satellites jointly serve networks of ground devices. Potential use cases are presented highlighting application requirements. We evaluate the average received power at the target locations using two coordination schemes and perform a statistical characterization of the power spillover on undesired locations. We also shed light on the performance of inter-satellite laser WPT for different operating distances and transmit-receive apertures of the peer satellites. Moreover, we explore the integration of metasurfaces on satellite apertures and ground networks to boost energy conversion efficiency, scalability, and beam management.  Finally, we outline relevant challenges and research directions towards implementing our vision. 
    \end{abstract}

    \section{Introduction}\label{sec:introduction}   
    Wireless power transfer (WPT) is a potential technology that can support the growing and heterogeneous energy demands of Internet of Things (IoT) networks. It relies on energy transmitters (ETs) to wirelessly recharge energy harvesting (EH) devices. By eliminating battery replacements/recharging, WPT can lower maintenance costs and e-waste, enable smaller devices, and support temporary or mission-specific deployments.
    
    Battlefield logistics, remote monitoring, and disaster relief may require rapid and infrastructure-free energy provisioning, which may go beyond the feasible limits of traditional WPT. Although unmanned aerial vehicle (UAV)-based ETs can support time-critical missions, their limited autonomy constrains operation in remote or rapidly evolving scenarios.

    Space-based WPT can extend energy provisioning beyond the power grid and into hard-to-reach areas. First proposed by Peter Glaser in 1968 for collecting solar energy in orbit and beaming it to Earth by microwaves, the concept has since motivated numerous research initiatives and demonstrations \cite{Rodenbeck.2021, Feng.2026}. Nevertheless, charging low-power IoT devices from space remains challenging because of long propagation distances and limited visibility windows, motivating architectures that combine energy contributions from multiple satellites.
    
    At the same time, rising demand for ubiquitous 6G connectivity and reduced launch costs are driving increasingly large satellite constellations. Leading companies such as SpaceX, SES, Eutelsat Group, Amazon, Boeing, Huawei, and Telesat plan to deploy tens of thousands of satellites \cite{Hraishawi.2023}. Their increasing density creates new opportunities for space-based WPT, as multiple satellites may be simultaneously visible and jointly charge ground devices.

    Space-based WPT spans inter-satellite and space-ground energy transfer \cite{Rodenbeck.2021}. Inter-satellite charging can use high-orbit platforms with longer sunlight exposure and larger energy reserves to supply lower-orbit satellites. The high line-of-sight (LoS) availability of inter-satellite links also makes laser-based WPT attractive because of its compact and lightweight hardware. In contrast, microwave WPT is better suited to space-ground charging, where strict LoS may not always be available. Existing space-ground research has largely targeted grid-scale energy transmission to Earth, while multi-satellite WPT for IoT remains largely unexplored beyond communication link-budget enhancement and single-user charging \cite{Xu.2023,Shehab.2026}. Dense LEO constellations are particularly attractive because their lower altitudes reduce path loss and delay compared with medium- and geostationary-orbit systems.

   Motivated by this opportunity, we propose a coordinated space-based WPT framework in which a dynamically selected set of visible or soon-to-be-visible satellites jointly charges low-power IoT devices. Coordination can be supervised by a ground network or enabled through direct inter-satellite links. While radio frequency (RF)-based space-to-ground WPT is envisioned as the primary service, inter-satellite WPT may provide complementary energy to sustain serving satellites. By aggregating multiple satellites, coordinated space-based WPT can offer higher beamforming gains and received power density, extended service windows through overlapping coverage, reduced aperture size and power requirements per satellite, and enhanced robustness against fading through spatial diversity. We make the following contributions:
   \begin{figure*}[t!]
    \centering
    \includegraphics[width=0.98\linewidth]{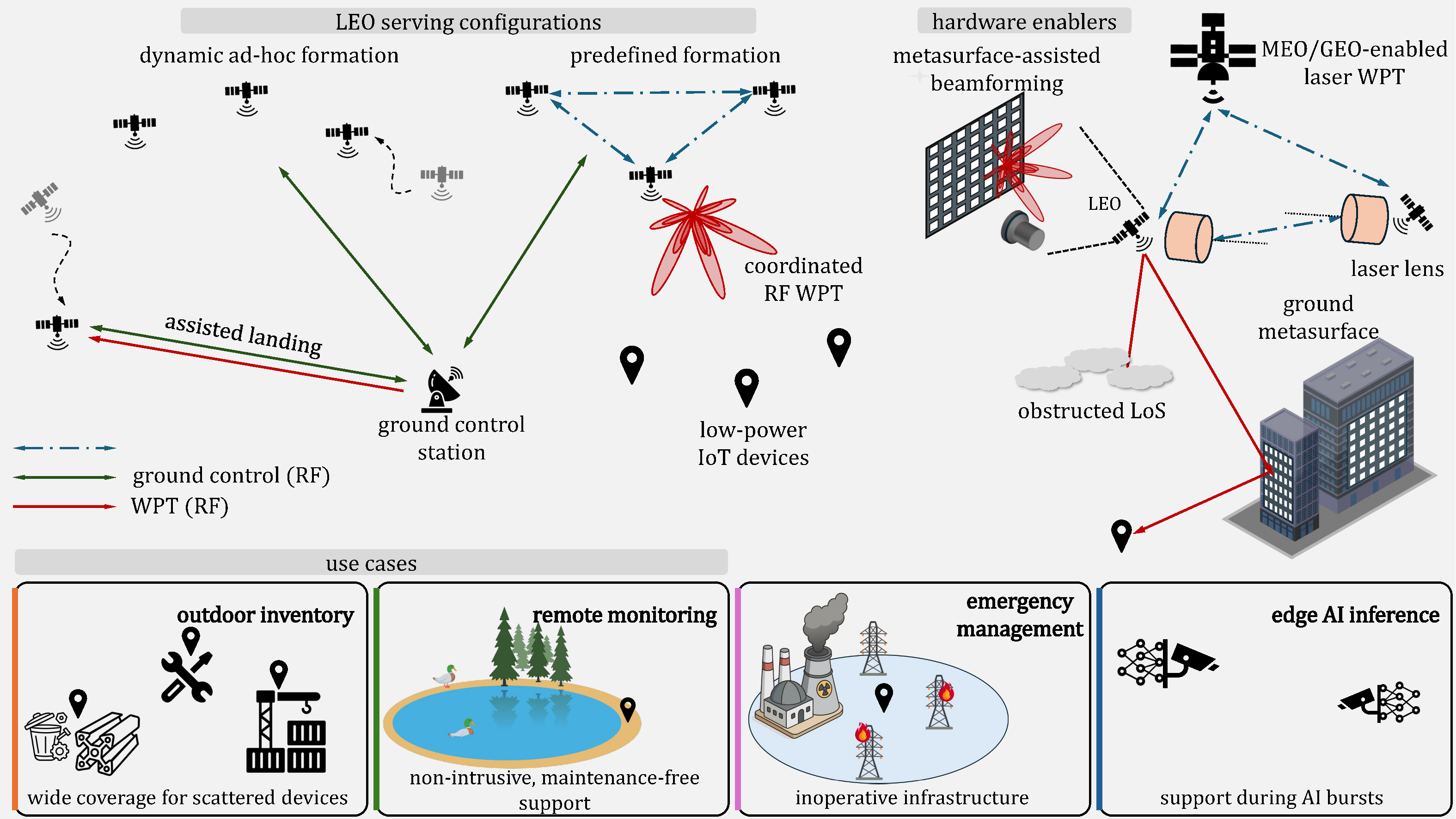}
    \caption{Illustration of the coordinated space-based WPT framework along with enabling technologies and potential use cases. Two LEO serving configurations are illustrated: i) dynamic ad-hoc formation, in which the serving set changes with satellite visibility and power contribution; and ii) predefined formation, in which the formation is maintained during orbit.}
    \label{fig:visionFigure}
\end{figure*}
   \begin{itemize}
       \item We overview the advantages, feasibility, and limitations of existing long-distance WPT technologies. 
       \item We characterize the power requirements of representative off-grid IoT use cases and discuss satellite-based WPT's supporting role.
       \item We envision coordinated multi-satellite WPT to charge low-power devices and examine received power under different coordination strategies. Moreover, we assess how link distance and aperture sizes affect inter-satellite laser-based WPT.
       \item We highlight relevant challenges throughout the article while pointing to potential research directions.
   \end{itemize}

\section{Long-distance WPT}\label{sec:longDistanceWPT}
Microwave-based WPT is recognized as a viable technology for long-distance wireless energy delivery \cite{ZHENG202417,Rodenbeck.2021}. The transmitter converts electrical power into electromagnetic radiation and directs it toward the intended region using directive antennas. At the receiver, a rectenna converts incident signal into electrical energy. The broadcast nature of microwave-based WPT enables the simultaneous charging of multiple ground users even in the absence of a direct LoS link. Fig.~\ref{fig:visionFigure} depicts different WPT scenarios. 

In space-to-ground WPT scenarios, the long transmission distances and propagation through the atmosphere introduce additional impairments beyond those encountered in terrestrial deployments. For instance, attenuation is proportional to cloud thickness, while rain and atmospheric gases contribute to absorption and scattering losses. Moreover, atmospheric turbulence induces scintillation, i.e., amplitude and phase fluctuations, whereas ionospheric effects lead to dispersion and polarization rotation. Consequently, the wireless channel exhibits a wide range of fading behaviors, with characteristics that vary significantly depending on LoS conditions and elevation angles. Beyond these propagation effects, hardware design, onboard power/processing constraints, and satellite geometry and operation also influence end-to-end power delivery. Given the severe free-space path loss over satellite-to-ground distances, these components must be carefully optimized/considered to mitigate diffraction and spillover losses, and hence ensure sufficient power delivery to the receiver. In this context, ground-supported metasurfaces can enlarge the aperture of the energy capturing system while enabling indirect LoS links.

Several experiments have validated the feasibility of long-distance WPT. Large-scale demonstrations, such as the Goldstone Deep Space Communications Complex experiment, achieved tens of kilowatts (kW) over kilometer (km)-scale distances with end-to-end efficiencies exceeding $50 \%$. Contemporary experiments in Japan, United States, China, and South Korea have demonstrated kW-level WPT over hundreds of meters. However, a clear dichotomy persists: high end-to-end efficiency is typically achieved in short-range links, whereas it severely degrades for longer distances. Recently, \cite{Biao.2021} reported a microwave-based WPT system achieving $1 \%$ end-to-end efficiency over $10$ km, highlighting that while km-scale WPT is technically feasible, it remains constrained by a low end-to-end efficiency despite using highly directive apertures and specialized RF-direct current (DC) conversion architectures. Moreover, a Japanese firm has demonstrated ground-to-ground and sky-to-ground WPT for $50$ m to $5$ km distances \cite{ssps}. 

To support long-distance WPT, several architectures have been proposed including the sandwich architecture, where photovoltaic layers harvest solar energy and covert it to RF power before transmitting via phased arrays. However, solar illumination is inherently non-uniform. Indeed, LEO platforms and CubeSats have limited surface area for solar cells, and nearly $40$\% of their orbit lies in Earth's shadow. Fortunately, higher orbit satellites have longer sunlight exposure and can generate more electricity. Their surplus electricity can be transferred to lower orbit satellites. In \cite{Mqaraqe}, satellite-based HotSpots are introduced, where higher orbit satellites transfer surplus energy to lower orbit satellites using optical WPT. Other demonstrations \cite{ZHENG202417},\cite{DEdoardo,Flopez} show the potential of laser-based WPT; however, LoS requirement and strong atmospheric sensitivity render it unsuitable for space-to-ground WPT.

Table~\ref{tab: micro_las} summarizes key differences between microwave- and laser-based WPT. Henceforth, we consider microwave-based WPT for space-to-ground charging and laser-based WPT for inter-satellite charging. 

\begin{table}[t!]
    \centering
    \caption{Microwave vs Laser WPT}
    \begin{tabular}{p{1.3cm}p{3cm}p{2.9cm}} \toprule
         & \ \ \textbf{Microwave}     & \ \  \textbf{Laser} \\
    \toprule
    Frequency     &  $0.3-300 $ GHz &  THz range \\ \hline \\ \vspace{-3 mm}
    Free space path loss    &\vspace{-3 mm} 
    Moderate & \vspace{-3 mm} High due to high frequency \\ \hline \\ \vspace{-3 mm}
    Beam    &\vspace{-3 mm} 
    Wider coverage  &\vspace{-3 mm} Coherent (i.e, narrow beam) \\ \hline \\ \vspace{-3 mm}
    Tx power &\vspace{-3 mm} W $\sim$ GW & \vspace{-3 mm} W $ \sim$ MW \\ \hline \\ \vspace{-3 mm}
    Receiver efficiency & \vspace{-3 mm} up to $ 70 \%$ &\vspace{-3 mm} $ 10 \sim 25 \%$ \\ \hline \\ \vspace{-3 mm}
    Transmit distance & \vspace{-3 mm} m $\sim 10^2$ km  & \vspace{-3 mm} m $\sim 10^3$ km \\ \hline \\ \vspace{-3 mm}
    Typical losses & \vspace{-3 mm} Shadowing, attenuation, and multipath fading &\vspace{-3 mm} Pointing loss due to mechanical jitter, attenuation, and atmospheric turbulence\\
    \bottomrule
    \end{tabular}
    \label{tab: micro_las}
\end{table}

\section{Use Cases}\label{sec:useCases}
Satellite-based WPT is most advantageous when devices are geographically dispersed, difficult to access, and local charging infrastructure is impractical or costly. From a system-design perspective, terrestrial-, UAV-, and satellite-based WPT occupy complementary operating regimes. Terrestrial and UAV-based ETs are generally preferable when relatively high received power is required and deployment is feasible. Meanwhile, satellite-based WPT becomes attractive when service availability and response time outweigh power delivery. Accordingly, we consider the following representative low-duty-cycle IoT use cases.

\begin{itemize}
    \item \textbf{Outdoor inventory}: Large industrial areas or remote storage yards may contain widely distributed sensing or tracking devices, making dense terrestrial ET deployment costly and frequent UAV-based charging inefficient.
    
    \item \textbf{Remote monitoring}: Wildlife and environmental monitoring in protected regions requires long-term operation with minimal intervention, while deploying terrestrial ET infrastructure may be impractical and/or environmentally intrusive.
    
    \item \textbf{Emergency management}: After disasters damage terrestrial communication and energy infrastructures, satellite-based WPT can charge sparse low-power emergency sensors, alarm tags, and backscatter devices.
    
    \item \textbf{Edge AI inference}: Remote IoT devices may occasionally perform energy-intensive local inference. Satellite-based WPT can support these bursts when battery replacement and persistent local charging infrastructure are impractical.
\end{itemize}

Satellite-to-ground wireless links suffer from severe propagation loss, so feasibility must be assessed against realistic device power budgets. In IoT nodes, the dominant energy consumers are typically the central processing unit (CPU), sensing module, and radio. However, their contribution depends on the application profile. For example, small-bit CPUs used for lightweight control tasks consume 5-10~$\mu$W in active mode, whereas high-performance processors required for AI tasks may demand more than 100~mW. The sensing module may include sensors with power usage ranging from sub-mW levels for simple tasks (e.g., temperature sensing) to over 10~mW for industrial measurements. Finally, the radio module enables data transmission, either via active transmission, typically consuming 1-10~mW, or using ultra-low-power backscatter communication consuming 1-10~$\mu$W \cite{3gppamb,sensor_values}, suitable for event-driven emergency applications. Note that for backscatter-based radios, only RF-based wireless energy sources are applicable. Table~\ref{tab:energy_per_use_case_duty} summarizes representative power consumption levels and activity duty cycles for the CPU, sensing module, and radio in the considered use cases. These values capture different operational profiles of low-duty-cycle IoT devices and are used to estimate the total DC average power consumption. The corresponding required RF power is then obtained assuming a $50\%$ conversion efficiency \cite{Rodenbeck.2021}. The required RF power levels are used as a reference in Section~\ref{sec:coordinatedWPT} to evaluate the theoretical feasibility of coordinated satellite-based WPT.

\begin{table}[t]
    \centering
    \caption{Typical power profiles and average DC and required RF power for each use case.}
    \label{tab:energy_per_use_case_duty}
    \setlength{\tabcolsep}{4pt}
    \renewcommand{\arraystretch}{1.15}
    \begin{tabular}{
        p{0.21\columnwidth}
        p{0.36\columnwidth}
        p{0.15\columnwidth}
        p{0.15\columnwidth}
    }
    \toprule
    \textbf{Application}
    & \textbf{Profile}
    & \textbf{DC power}
    & \textbf{RF power} \\
    \midrule

    \makecell[l]{Outdoor\\inventory}
    & \makecell[l]{
        CPU: $5\%$, $10~\mathrm{mW}$\\
        Sensor: $10\%$, $0.5~\mathrm{mW}$\\
        Radio: $<1\%$, $5~\mathrm{mW}$
      }
    & $0.60~\mathrm{mW}$
    & $1.20~\mathrm{mW}$ \\

    \midrule
    Remote monitoring
    & \makecell[l]{
        CPU: $10\%$, $10~\mathrm{mW}$\\
        Sensors: $20\%$, $1~\mathrm{mW}$\\
        Radio: $<1\%$, $5~\mathrm{mW}$
      }
    & $1.25~\mathrm{mW}$
    & $2.50~\mathrm{mW}$ \\

    \midrule
    Emergency
    & \makecell[l]{
        CPU: event-driven\\
        Sensors: --\\
        Radio: $<1\%$, $5~\mu\mathrm{W}$
      }
    & --
    & -- \\

    \midrule
    AI application
    & \makecell[l]{
        CPU: $2\%$, $100~\mathrm{mW}$\\
        Sensors: $4\%$, $1~\mathrm{mW}$\\
        Radio: $<1\%$, $5~\mathrm{mW}$
      }
    & $2.09~\mathrm{mW}$
    & $4.18~\mathrm{mW}$ \\

    \bottomrule
    \end{tabular}
\end{table}

\section{Coordinated space-based WPT}\label{sec:coordinatedWPT}
Fig.~\ref{fig:visionFigure} illustrates two LEO formations. In one, satellites belong with a predefined swarm that maintains its formation while orbiting. In the other, an ad-hoc serving group is formed from satellites visible to the ground devices. 

On Earth, a network of ground control stations (GCSs) coordinates with the satellites to serve devices' charging requests. Given the rapid movement of LEO satellites, the GCSs update the serving set in real time, adding satellites as they enter the visibility window and removing those about to leave or contribute negligibly. This minimizes charging interruptions and unnecessary satellite energy expenditure.

Constructive interference requires sub-wavelength phase coherence and precise time-frequency alignment to compensate for altitude jitter, oscillator drift, and Doppler-induced phase shifts. Such synchronization may be achieved via closed-loop techniques, where each satellite continuously measures and corrects its phase and propagation delay through a feedback link to a designated GCS. Alternatively, open-loop synchronization can be employed, where satellites exchange timing and frequency information directly through ISL. Closed-loop algorithms are well-suited to ad-hoc swarms, which are more dependent on a GCS, whereas open-loop synchronization is more attractive for clustered formations with lower ISL delays \cite{Marrero.2022}. ISL modules can be also utilized for for inter-satellite ranging and estimation of relative positioning which are supplementary functions to correct small location drifts.

\begin{figure}[t]
    \centering
    \includegraphics[width=\linewidth]{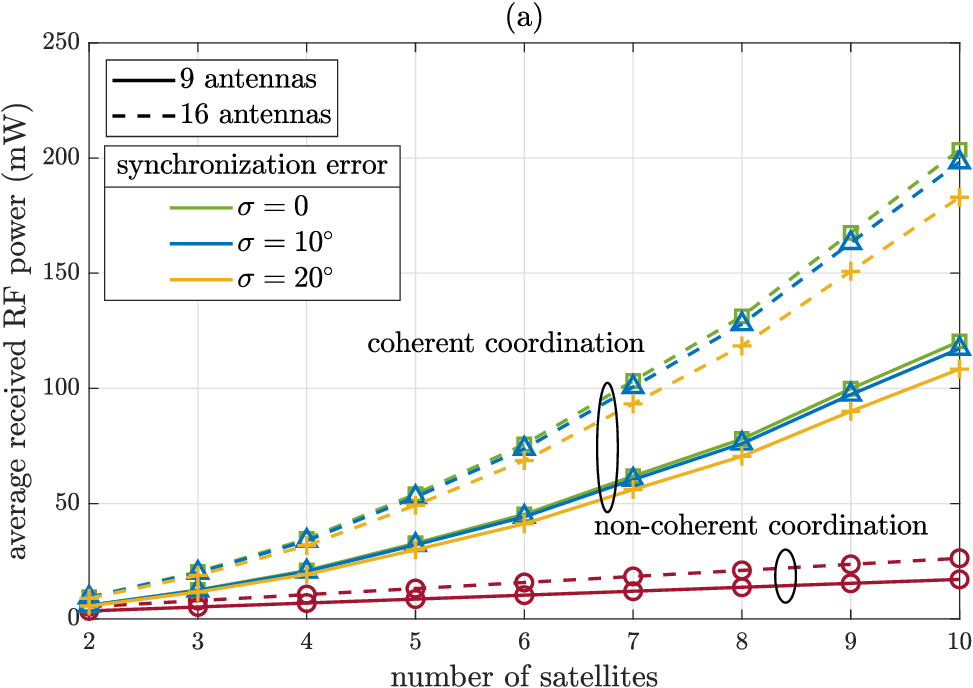}\\
    \vspace{0.5em}
    \includegraphics[width=\linewidth]{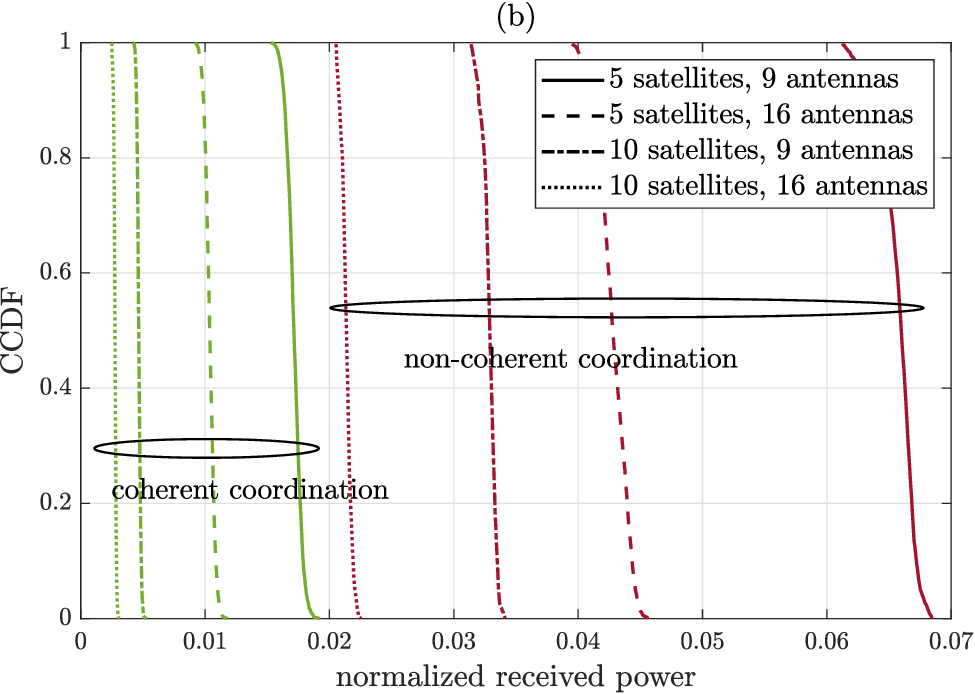}
    \caption{Distributed WPT with the satellites deployed at the vertices of a regular polygon with circumradius $10~$km and orbiting at an altitude of $500~$km, thus resembling a mesh configuration. Each satellite is equipped with a uniform rectangular array (pointing towards the nadir) and transmits at an operating frequency of $12~$GHz with a maximum power budget of $1000~$W. Results were obtained by maximizing the minimum received power across the devices assuming free-space LoS satellite-to-ground channels. Five single-antenna devices are uniformly distributed in a circular area of radius $2~$km. a) Average received power vs the number of visible satellites. b) Complementary cumulative distribution function (CCDF) of the average received power in locations excluding those occupied by the target devices and normalized with respect to both the number of visible satellites and the average received power at the devices. The residual phase error at each satellite is modeled as an independent zero-mean Gaussian random variable with standard deviation $\sigma$. Both results correspond to single network snapshot optimizations without accounting for the temporal evolution of the system.}
    \label{fig:rfwetSimulation}
\end{figure}

Fig.~\ref{fig:rfwetSimulation}a illustrates how the average received RF power increases as the number of serving satellites grows. Two transmission strategies are considered: i) coherent coordination, where all satellites form a distributed antenna system and transmit common phase-aligned signals; and ii) non-coherent coordination, where the satellites serve the devices jointly but transmit independent signals without phase synchronization. Each satellite-to-device link is modeled as a narrowband line-of-sight channel. For every device deployment, the array precoders are jointly optimized under an individual power budget per satellite to maximize the power delivered to the worst-served device. Under coherent transmission, the received power scales quadratically with the number of satellites and increases further with more antennas. In contrast, non-coherent transmission scales linearly and benefits only modestly from additional antennas. Fig.~\ref{fig:rfwetSimulation}b further shows that coherent coordination concentrates energy more strongly at the target devices, while increasing the number of antennas and satellites reduces spillover in both strategies.

The received power in Fig.~\ref{fig:rfwetSimulation} ranges from sub-mW levels with a few non-coherently coordinated satellites to tens of mW for coherent coordination with larger arrays and more satellites. For comparison, Table~\ref{tab:energy_per_use_case_duty} reports the required RF power levels, which range from a few $\mu$W for event-driven tags up to a few mW for edge-AI nodes. Therefore, configurations delivering between $1.20$ and $2.50~\mathrm{mW}$ can theoretically support outdoor inventory, those delivering at least $2.50~\mathrm{mW}$ can also support remote monitoring, and those exceeding $4.18~\mathrm{mW}$ can satisfy the average power requirements of all three applications.

Note that the synchronization requirements differ significantly between the two strategies. While non-coherent coordination operates with conventional per-link timing and frequency synchronization and does not require inter-satellite phase alignment, the coherent coordination strategy imposes substantially more stringent synchronization requirements, as shown in Fig.~\ref{fig:rfwetSimulation}a. In particular, at $12~$GHz, achieving near-constructive addition requires sub-nanosecond timing accuracy (e.g., to keep residual phase errors below approximately $20^\circ$) and very tight frequency synchronization. The impact of imperfect synchronization is illustrated in Fig.~\ref{fig:rfwetSimulation}a, where the received power achieved by coherent coordination decreases as the standard deviation of the phase errors increases. Consequently, practical systems deviate from the ideal coherent-coordination benchmark depending on the residual synchronization errors and may therefore operate in a partially coherent or even non-coherent regime.

Sustaining long-term satellite operation becomes particularly challenging when supporting WPT services, which require handling higher instantaneous power levels than in traditional communication services. These additional requirements motivate the development of complementary energy supply mechanisms to sustain satellite operation. In particular, inter-satellite laser-based WPT may become appealing to harness energy from satellites with better sun exposure. However, reaping the full benefits of this technology requires precise transmitter and receiver alignment. 

\begin{figure}
    \centering
    \includegraphics[width=\linewidth]{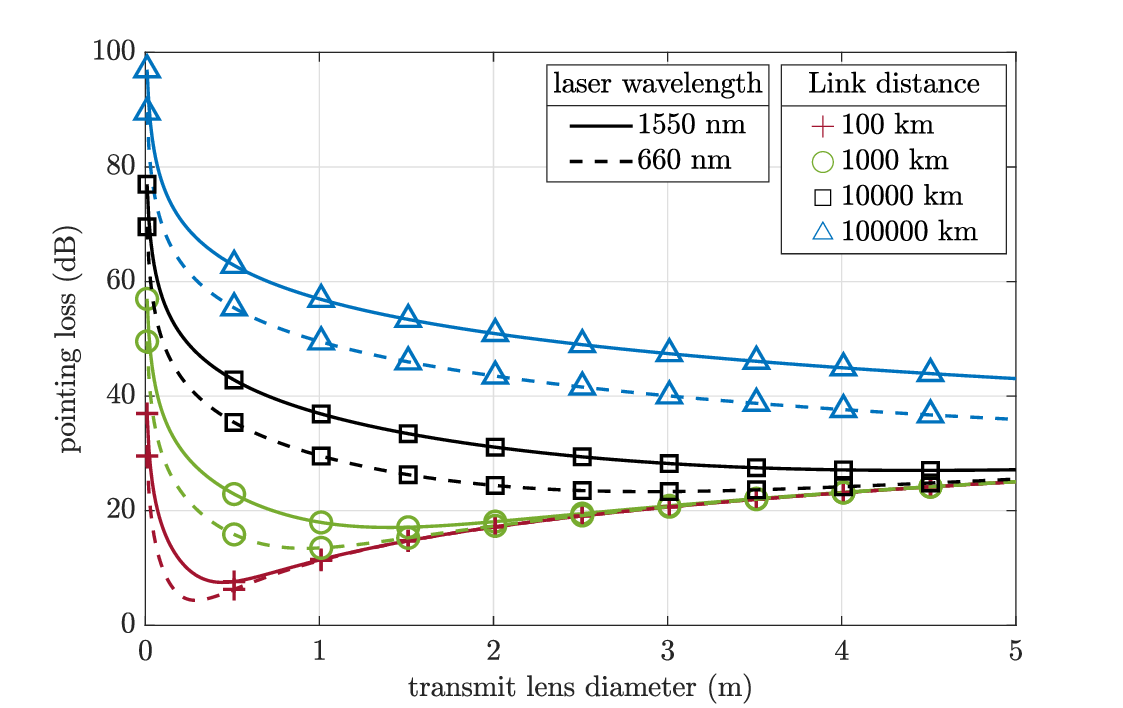}
    \caption{Pointing loss with respect to the aperture of the transmitter. The variations in the pointing loss are captured through pointing error, which is assumed to follow power-law distribution with parameter $6.7$.}
    \label{fig:LwetSimulation}
\end{figure}

Fig.~\ref{fig:LwetSimulation} illustrates the effects of laser beam misalignment, transmission distance, transmitter aperture, and optical wavelength on the pointing loss in a point-to-point inter-satellite WPT scenario. The pointing loss first decreases and then increases with transmit aperture for every set of parameters. This is due to the fact that a small transmit aperture results in a dense laser beam and thus even a small misalignment can deteriorate the received irradiance levels significantly. On the other hand, a larger transmit aperture produces a wider, less dense beam, reducing irradiance at the receiver and increasing pointing loss. Furthermore, we observe that the pointing loss increases with link distance. This results from beam spreading, which reduces the received irradiance. Such behavior reveals an optimal transmit aperture that minimizes pointing loss, motivating adaptive techniques such as variable beam expansion.

At lower altitudes, atmospheric drag induces orbit decay, which must be corrected to maintain precise pointing and to avoid collisions with nearby satellites and debris. These constraints demand onboard power-management strategies to prevent blackouts such as load-shedding algorithms that deactivate non-critical subsystems during power scarcity. Furthermore, thermal control via active cooling or heaters can incur substantial power draw, especially in shadowed segments, further straining on-board batteries. Notably, optimization of the solar panels for efficient harvesting of both laser and sunlight is of paramount importance to boost end-to-end conversion efficiency. Finally, during emergency scenarios, GCSs can act as external energy suppliers to sustain satellite operations or assist in controlled de-orbiting maneuvers. 

\section{Metasurface-assisted WPT}\label{sec:metasurfaces}
Metasurfaces are planar structures comprising sub-wavelength elements that offer control over electromagnetic wave propagation. This section highlights two applications of metasurface-assisted WPT: i) integrating metasurfaces into satellite-based ETs for beamforming with reduced complexity; and ii) deploying them near receivers to overcome non-LoS conditions and enhance EH efficiency as shown in Fig.~\ref{fig:visionFigure}. These applications are motivated by satellite-specific constraints such as long propagation distances, limited size/weight/power budgets, and dynamic satellite-ground geometry.

Metasurfaces complement coordinated satellite-based WPT. For coherent coordination, metasurface-assisted beam control provides additional spatial degrees of freedom for concentrating energy, while inter-satellite phase and timing synchronization remains necessary. For non-coherent coordination, metasurfaces can improve individual-link directivity or provide additional propagation paths without phase coherence among serving satellites.

Reconfigurable metasurfaces, such as dynamic metasurface antennas (DMAs) and reconfigurable holographic surfaces (RHSs), offer low-power, scalable beamforming. For instance, DMAs reduce hardware complexity by using a few RF chains to feed multiple elements via waveguides. Nevertheless, employing DMAs and RHSs in space-based WPT requires careful consideration of high-power handling, since unit cells must tolerate intense surface currents and may suffer from dielectric heating or breakdown. Alternatively, parabolic reflectarray technology \cite{silva2019metasurface} enables beam collimation and focusing without bulky feed networks or active phase shifters.

For on-orbit deployment, the large effective aperture compensating for severe space-to-ground path loss must be balanced against satellite mass and launch-volume constraints, motivating planar, foldable, or deployable surfaces. Deployment tolerances and structural deformation can perturb element phases, while attitude errors can cause beam-pointing mismatch, requiring calibration and adaptive phase compensation. Because satellite-ground geometry changes during LEO passes, the metasurface must dynamically adapt as multiple satellites enter and exit its coverage zone. In coherent operation, local beam control must coexist with inter-satellite synchronization, whereas non-coherent operation can enhance per-satellite focusing without phase alignment. Joint optimization of metasurface states, satellite selection, and charging intervals remains an important research direction.

The second application is deploying reflective surfaces, such as reconfigurable intelligent surfaces (RISs), near receivers to capture and redirect otherwise lost energy, creating virtual LoS links in blocked environments. Recent work on multi-RIS-assisted space-air-ground integrated networks has investigated RISs for adaptive propagation control \cite{Feng.2026}. However, limited phase resolution, nonlinearity in tunable elements, and mutual coupling can reduce beamforming accuracy and efficiency, requiring careful design and calibration.

\section{Conclusion \& Outlook}\label{sec:conclusionsAndOutlook}
The surge in satellite communications creates new opportunities for energy provision in remote off-grid areas, disaster scenarios, and military applications. This article reviewed the potential and challenges of RF- and laser-based space WPT and introduced coordinated satellite charging for low-power ground devices. We examined how satellite coordination, array size, and the number of satellites affect received power; evaluated pointing errors in inter-satellite laser WPT; and discussed metasurfaces for reducing ET complexity, creating virtual LoS links, and improving power reception. Representative applications and their service requirements were also presented.

We next summarize key challenges and research directions for coordinated charging from space.

\textbf{Distributed beamforming management:} LEO motion continuously changes propagation delays, Doppler shifts, and serving-set geometry. If not accurately tracked and compensated, these variations cause synchronization and beam-steering errors, reducing gain or creating unintended nulls \cite{Xu.2023,Hraishawi.2023,Shehab.2026}. The problem is more severe at high frequencies, where sub-ns timing errors can produce substantial phase misalignment. Beyond real-time re-synchronization, robust beam management should accommodate satellites joining or leaving the serving set through redundancy, adaptive re-optimization, and geometry-aided predictive beamforming. Optimization and reinforcement-learning strategies at the GCS may further improve system-wide resilience.

\textbf{Satellite autonomy:} WPT power allocation must not compromise the satellite's primary functions, particularly during eclipse periods. High-power WPT may require larger apertures, RF amplification, energy storage, and thermal management, increasing mass and hardware complexity \cite{Rodenbeck.2021}. Practical implementations should therefore reuse communication apertures and RF hardware and adapt WPT to onboard energy and thermal conditions. Energy-aware satellite selection, adaptive duty-cycling, load distribution, and inter-satellite WPT from higher-orbit platforms can further reduce the burden on individual satellites.

\textbf{Safety concerns:} High-power inter-orbital laser beams may propagate to Earth's surface due to pointing errors, where their intensity could be hazardous. Therefore, exposure to laser transmissions must be minimized. A possible solution is a feedback mechanism that informs the transmitter whether the beam reaches the intended receiver. Based on this feedback, the transmitter can interrupt transmission when alignment is lost and resume it after re-establishing the LoS link between transmitter and receiver. Further studies on high-power transfer to ground, associated safety constraints, and feasible implementation zones on Earth are also necessary during standardization.

\textbf{Positioning and tracking for laser-based WPT:} A key challenge in laser-based WPT is the accurate positioning and tracking of the non-static target satellite for reliable beam alignment. Although two-line element (TLE) data provide satellite position information, the relatively large temporal gaps between successive TLEs may limit the availability of up-to-date, arcsecond-level position information for precise laser beam pointing. A possible solution is to employ microwave-based positioning and tracking as an assisting technology to obtain more timely and accurate target-satellite positions. This consequently introduces a resource-allocation problem in which the available energy must be appropriately divided between microwave-based positioning and laser-based WPT.

\textbf{Variable conversion efficiency in laser-based WPT:} The optical-to-electrical conversion efficiency of laser-based WPT is affected by the thermal conditions of the photovoltaic receiver \cite{DEdoardo,ZHENG202417}, which can vary during the power-transfer period due to constant motion. Consequently, a constant transmit power may be suboptimal as the conversion efficiency changes with the receiver temperature. A possible solution is therefore to dynamically adapt the laser beam intensity according to the instantaneous thermal conditions, with the objective of maximizing the overall energy-transfer efficiency. This can also provide insights into the achievable efficiency limits and practical feasibility of laser-based WPT for ISLs.

\textbf{Channel acquisition and beam-control overhead:} Coordinated satellite-based WPT requires sufficiently accurate channel, phase, timing, and satellite-device geometry information. Rapidly varying LEO geometry, Doppler shifts, and propagation delays limit the validity of quasi-static channel estimates, while perfect instantaneous CSI may overestimate coherent beamforming gains. Metasurfaces further increase configuration overhead due to their large number of controllable elements. Possible solutions include geometry- and ephemeris-aided channel/beam tracking, partial or statistical CSI, adaptive pilot/feedback scheduling, and multi-timescale beam control to compensate for residual channel and synchronization errors.

\bibliographystyle{IEEEtran}
\bibliography{IEEEabrv,referencesShort}

\vspace{-1em}
\section*{Biographies}
\small
\noindent\textbf{Osmel M. Rosabal} [S'21 M'25] is a postdoctoral researcher, focused on RF localization and sensing, at the Centre for Wireless Communications (CWC), University of Oulu, Finland.

\noindent\textbf{Amirhossein Azarbahram} [S] is a postdoctoral researcher at the CWC, University of Oulu, Finland, focused on AI-enabled wireless communications and RF sensing.

\noindent\textbf{Mateen Ashraf} is a postdoctoral researcher at the CWC, Oulu, Finland, focused on long-distance wireless charging protocols.

\noindent\textbf{Mohammad Shehab} is an Assistant Professor of Networks with the German University in Cairo, Egypt. 

\noindent\textbf{Abdul Basit Khattak} [S] is a Doctoral researcher at the CWC, University of Oulu, Finland, focused on RF wireless power transfer and sensing.

\noindent\textbf{Onel L. A. L\'opez} [S'17, M'20, SM'24] is an Associate Professor in sustainable wireless communications engineering at the CWC, Oulu, Finland.

\noindent\textbf{Mohamed-Slim Alouini} [F'09] is a Distinguished Professor of Electrical Engineering at KAUST, Saudi Arabia. 
\end{document}